\documentclass[aps,pre,twocolumn,superscriptaddress,showpacs,longbibliography]{revtex4-1}
\usepackage[dvips]{graphicx}
\usepackage{amsmath}

\usepackage{graphicx}
\usepackage{dcolumn}
\usepackage{bm}

\usepackage[utf8]{inputenc}
\usepackage[T1]{fontenc}
\usepackage{mathptmx,float}
\usepackage{textcomp}
\usepackage{xcolor}

\begin{document}

\title{Impact of field heterogeneity on the dynamics of the forced Kuramoto model}

\author{S. Yoon}
\affiliation{Department of Physics $\&$ I3N, University of Aveiro, 3810-193 Aveiro, Portugal}

\author{E. A. P. Wright}
\email[Corresponding author: ]{wrighteap@ua.pt}
\affiliation{Department of Physics $\&$ I3N, University of Aveiro, 3810-193 Aveiro, Portugal}

\author{J. F. F. Mendes}
\affiliation{Department of Physics $\&$ I3N, University of Aveiro, 3810-193 Aveiro, Portugal}

\author{A. V. Goltsev}
\affiliation{Department of Physics $\&$ I3N, University of Aveiro, 3810-193 Aveiro, Portugal}
\affiliation{A.F. Ioffe Physico-Technical Institute, 194021 St. Petersburg, Russia}

\date{\today}

\begin{abstract}
  We studied the impact of field heterogeneity on entrainment in a system of uniformly interacting phase oscillators. Field heterogeneity is shown to induce dynamical heterogeneity in the system. In effect, the heterogeneous field partitions the system into interacting groups of oscillators that feel the same local field strength and phase. Based on numerical and analytical analysis of the explicit dynamical equations derived from the periodically forced Kuramoto model, we found that the heterogeneous field can disrupt entrainment at different field frequencies when compared to the homogeneous field. This transition occurs when the phase- and frequency-locked synchronization between groups of oscillators is broken at a critical field frequency, causing each group to enter a new dynamical state (disrupted state). Strikingly, it is shown that disrupted dynamics can differ between groups.



\end{abstract}

\maketitle

\section{Introduction}
\label{sec:Introdution}

The emergence of self-sustained rhythms - synchronization - and externally driven rhythms - entrainment - is ubiquitous in nature, and drives considerable scientific inquiry across theoretical and applied disciplines, from biology to sociology. Entrainment has been observed in biological systems, where light-dark cycles entrain the circadian rhythms of organisms \cite{Silver2018,Michel2020}, seasonal cycles entrain physiological and behavioral changes in ecological networks (e.g. migration and adaption to different seasons), lunar cycles entrain physiological and behavioral modifications including metamorphosis \cite{Brugler_2018}, and periodic external sensory stimuli entrain brain waves (brainwave entrainment) \cite{Namazi2015,Mosbacher2020}, among others. Likewise, the operation of technological networks (transportation, power consumption, {\it etc.}) is entrained to light-dark cycles, and periodic social events such as elections entrain the activity of social and economic networks.

Understanding the mechanisms of entrainment is an open problem in modern interdisciplinary investigations of complex systems. In many real systems, a periodic external field acts only on a subset of the system, or acts on different subsets with different phases and strengths. Consider, for example, that in many moth species, periodic environmental cues only entrain sex pheromone biosynthesis in the females, driving males to mate, and therefore impacting the entire species \cite{Silvegren2005}. Further examples of how field heterogeneity is determinant to system-wide behavior can be found in the suprachiasmatic nucleus (SCN) of mammals, which coordinates periodic physiological and behavioural changes over 24-hour cycles \cite{Silver2018,Michel2020,Evans2016}. In the SCN, photic input from the retina entrains a subset of oscillators (neurons with periodic genetic and electrical activity) in the ventral ("core") region \cite{Meijer1986}. In turn, these oscillators entrain the remaining cells in the SCN to light-dark cycles. Other subsets of oscillators within the SCN have been implicated in non-photic regulation of circadian rhythms \cite{Yuan2018}. As an example of a non-photic input, consider changes in feeding schedule, which have been shown to shift and/or entrain gene expression in the SCN under caloric restriction \cite{Mendoza2005,Mendoza2007}. In this case, SCN activity demonstrates simultaneous entrainment to photic and non-photic cues over 24-hour cycles. 

Despite the evident importance of field heterogeneity for the functioning of real systems, the impact of field heterogeneity on entrainment has not been studied. In this paper, we consider entrainment by a heterogeneous external field in a synchronized system of coupled phase oscillators. Our analysis is based on the Kuramoto model, a representative model of synchronization \cite{Rodrigues2016,Arenas2008,Acebron2005,Strogatz2000}. Based on this model, we study how the interplay between field frequency and heterogeneity impacts entrainment, how entrainment is disrupted, and the dynamics of the disrupted states.

In Section~\ref{sec:kmhf}, we introduce the equations for the Kuramoto model in a heterogeneous field. In Section~\ref{sec:results}, we present a numerical analysis of the model for a field acting on two subsets of oscillators with distinct phases and strengths. In particular, we consider the impact of exposing a fraction of oscillators to the field, of varying the strength of the field on one subset while the other remains constant, and of varying the difference (shift) between field phases. In Section~\ref{sec:dynamics}, we discuss the dynamics of disrupted states from the point of view of an observer in the laboratory frame of reference. Finally, we discuss our findings and summarize our main results in Section \ref{sec:discussion}.

\section{Kuramoto model in a heterogeneous field}
\label{sec:kmhf}

Consider the Kuramoto model of \(N\) heterogeneous phase oscillators with all-to-all interactions, where each oscillator $i=1,2, \dots N$ has a natural frequency $\omega_i$, and natural frequencies are distributed according to some density function  $g(\omega)$. In a periodic external field with frequency $\sigma$, local phase $\sigma t + \phi_i$ and local strength $F_i$, the phase $\theta_i$ of each oscillator is determined by the following dynamical equation
\begin{equation}
\frac{d \theta_i}{dt}=\omega_i + \frac{K}{N}\sum_{j=1}^N\sin(\theta_j-\theta_i)+ F_i \sin(\sigma t + \phi_i-\theta_i),
\label{eq:km}
\end{equation}
where $K$ is the coupling constant.

Let us now consider the case where a heterogeneous external field acts with distinct strength and phase on different oscillators, dividing oscillators into groups that feel the same strength and phase. Formally, $M$ groups of $N_m$ oscillators are exposed to field phase $\sigma t + \phi_m$ and strength $F_m$, for $m=1,2, \dots, M$. In this paper, we label each group $G_m$, and consider the case where every group is large ($N_m\gg1$). In addition, we also assume that oscillators are distributed randomly over all groups. In other words, there are no correlations between the natural frequencies of oscillators in group $G_m$ and phase $\phi_m$ or strength $F_m$. For every group $G_m$, we then introduce the complex order parameter,


\begin{equation}
z_m=\rho_m e^{i\psi_m}\equiv \frac{1}{N_m}\sum_{j \in G_m} e^{i\theta_j},
\label{eq:zm}
\end{equation}
to characterize the state of  $G_m$. The amplitude $\rho_m$ characterizes the phase coherence between oscillators in $G_m$, and varies between $0$ and $1$. When $\rho_m=0$, oscillators within $G_m$ are in an asynchronous state, while $\rho_m=1$ corresponds to a completely synchronized state. The group phase $\psi_m$ characterizes the predominant direction of the oscillators. The global complex order parameter $Z$ for the entire system of oscillators may then be written as a sum of group order parameters $z_m$,
\begin{equation}
Z=\frac{1}{N}\sum_{j } e^{i\theta_j} = \sum_{m=1}^M f_m z_m,
\label{eq:zm}
\end{equation}
where $f_m \equiv N_m /N$ is the fraction of oscillators within group $G_m$, and $\sum_{m=1}^{M} f_m= 1$. Looking at Eq.~(\ref{eq:zm}), we see that the overall state of the system depends both on the state of each group and on the fraction of oscillators it contains.

Finally, let us assume that the natural frequencies of oscillators are distributed according to the Lorentz distribution function,
\begin{equation}
g(\omega)=\frac{\Delta}{\pi[(\omega -\overline{\omega})^2 + \Delta^2]},
\label{eq:df}
\end{equation}
where $\overline{\omega}$ is the average value of the natural frequencies and $\Delta$ is the spread (or full width at half maximum). Given that $N_m \gg1$ and the absence of correlations between natural frequencies and the field amplitude $F_m$ or phase $\phi_m$, the natural frequencies of oscillators within each group also are distributed according to the Lorentz distribution in Eq.~(\ref{eq:df}). This assumption allows us to employ the approach proposed by \cite{Ott2008,Ott2009}, and recently employed by Restrepo \emph{et al} \cite{Restrepo2019}, among others, to derive the following explicit self-consistent dynamical equation for $z_m$

\begin{eqnarray}
{\frac{d z_m}{d t}}=&& \frac{1}{2} \Bigl[K \sum_{n=1}^M f_n z_n + F_m e^{i\phi_m} -(K \sum_{n=1}^M f_n z_{n}^{*} + F_m e^{-i\phi_m}) z_{m}^2 \Bigr]
\nonumber\\
&& - \Bigl[\Delta + i\Omega \Bigr]z_m,
\label{eq: zm2}
\end{eqnarray}
in a frame of reference rotating at the field frequency $\sigma$, such that
\begin{equation}
\Omega \equiv \sigma - \overline{\omega},
\label{eq: omega}
\end{equation}
is the detuning parameter.

The real and imaginary parts of Eq.~(\ref{eq: zm2}) (multiplied by $e^{-i\psi_m}$) describe the dynamical evolution of the group amplitude $\rho_m$ and the group phase $\psi_m$,

\begin{eqnarray}
{\frac{d \rho_m}{d t}}=&& - \rho_m \Delta {+} \frac{1}{2} F_m (1{-}\rho_{m}^2) \cos(\phi_m {-}\psi_m) \nonumber\\
&&  + \frac{1}{2} K(1{-}\rho_{m}^2) \sum_{n =1}^M f_n \rho_n \cos(\psi_n -\psi_m),
\label{eq:rho_m} \\
{\frac{d \psi_m}{d t}}=&&-\Omega + F_m  \frac{(1+\rho_{m}^2)}{2\rho_m} \sin(\phi_m-\psi_m)\nonumber\\
&& +K\frac{(1+\rho_{m}^2)}{2\rho_m} \sum_{n=1}^M f_n \rho_n \sin(\psi_n -\psi_m).
\label{eq:psi_m}
\end{eqnarray}

Thus, in the thermodynamic limit, if a heterogeneous external field partitions a system of Kuramoto oscillators into large groups, each exposed to a distinct field phase and strength, the corresponding set of $N$ dynamical equations in Eq.~(\ref{eq:km}) is reduced to the set of $2M$ explicit equations for the group amplitude $\rho_m$ and the group phase $\psi_m$, presented in Eqs.~(\ref{eq:rho_m}) and (\ref{eq:psi_m}), respectively. This reduction remarkably simplifies the study of entrainment in systems of phase oscillators exposed to a periodic heterogeneous field, under the assumption that the natural frequencies $\omega_i$ in Eq.~(\ref{eq:km}) are distributed according to the Lorentz distribution function $g(\omega)$ in Eq.~(\ref{eq:df}).

\section{The impact of field heterogeneity}
\label{sec:results}

In general, an external field may act heterogeneously on any of $M$ groups of oscillators. However, specific aspects of real systems may be captured by a simplified model with two groups ($M=2$). For example, if we consider that only a fraction of oscillators (neurons) in the SCN receives photic input, we may seek to understand how the fraction of field-exposed oscillators affects entrainment in the absence of further inputs. Likewise, we may also seek to understand the extent to which entrainment is possible in the SCN if two interacting groups of oscillators are subject to inputs with equal periods but differing strengths and/or with a time delay (or phase shift) between them. These aspects of field heterogeneity are depicted schematically in Fig.~\ref{fig:figure_1} below, and are the main focus of this section.

\begin{figure}[htp]
\centering
\includegraphics{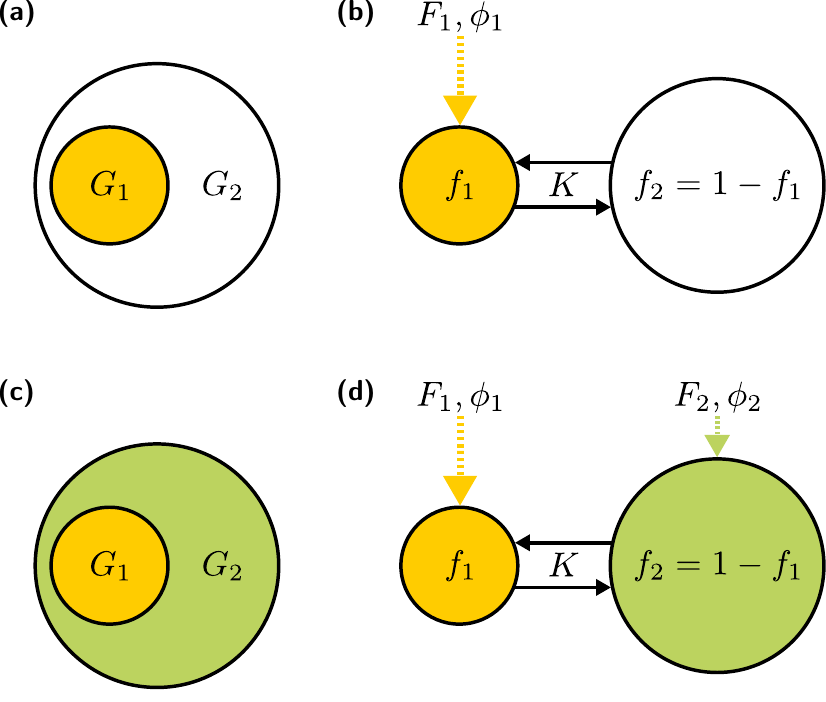}
	\caption{\label{fig:figure_1} Schematic depiction of a system of Kuramoto oscillators in a heterogeneous external field rotating with frequency \(\sigma\). Panel (a) depicts the simplest example of field heterogeneity, where only a subset of oscillators \(G_1\) is exposed to the field. Panel (b) shows an equivalent picture of the system as groups of interacting oscillators, where $K$ is the coupling strength of all oscillators: group \(G_1\) comprises the fraction $f_1$ of oscillators exposed to field strength \(F_1\) and phase \(\phi_1\), and group \(G_2\) comprises the fraction \(f_2 = 1 - f_1\) of oscillators that are not exposed to the field. Panels (c) and (d) depict a more general case, where both \(G_1\) and \(G_2\) are exposed to distinct, arbitrary field strengths and phases. Full arrows represent intra-group interactions, and dashed arrows represent field-group interactions.}
\end{figure}

For the cases depicted in Fig.~\ref{fig:figure_1}, where a periodic external field with frequency $\sigma$ acts with strength $F_1$ and phase $\phi_1$ on group of oscillators $G_1$, and strength $F_2$ and phase $\phi_2$ on group of oscillators $G_2$, the general set of Eqs.~(\ref{eq:rho_m}) and (\ref{eq:psi_m}) is reduced to

\begin{eqnarray}
{\frac{d \rho_1}{d t}}=&& - \rho_1 \Delta + \frac{K}{2} f_1 \rho_1 (1{-}\rho_{1}^2) {+} \frac{F_1}{2}  (1{-}\rho_{1}^2) \cos(\phi_1 {-}\psi_1) \nonumber\\
&&   + \frac{K}{2} f_2 \rho_2 (1{-}\rho_{1}^2) \cos(\psi_2 -\psi_1),
\label{eq: rho1} \\
{\frac{d \rho_2}{d t}}=&& - \rho_2 \Delta + \frac{K}{2}  f_2 \rho_2 (1{-}\rho_{2}^2) {+} \frac{F_2}{2} (1{-}\rho_{2}^2) \cos(\phi_2 {-}\psi_2) \nonumber\\
&&   + \frac{K}{2} f_1 \rho_1 (1{-}\rho_{2}^2)  \cos(\psi_1 -\psi_2),
\label{eq: rho2} \\
{\frac{d \psi_1}{d t}}=&&-\Omega +  F_1  \frac{(1+\rho_{1}^2)}{2\rho_1} \sin(\phi_1-\psi_1) \nonumber\\
&& + K \frac{(1+\rho_{1}^2)}{2\rho_1} f_2 \rho_2  \sin(\psi_2 -\psi_1),
\label{eq: psi1}\\
{\frac{d \psi_2}{d t}}=&&-\Omega + F_2  \frac{(1+\rho_{2}^2)}{2\rho_2} \sin(\phi_2-\psi_2) \nonumber\\
&& +K  \frac{(1+\rho_{2}^2)}{2 \rho_2} f_1 \rho_1 \sin(\psi_1 -\psi_2),
\label{eq: psi2}
\end{eqnarray}

\noindent
where $f_1+f_2=1$. For simplicity, we take $\Delta=1$ to be the frequency unit, and $\overline{\omega}=0$ to be the average natural frequency, so that the detuning parameter $\Omega$ is equal to the field frequency $\sigma$ (see Eq.~(\ref{eq: omega})). Through numerical analysis of Eqs.~(\ref{eq: rho1})--(\ref{eq: psi2}), we study the impact of the fraction of field-exposed oscillators $f_1$ (at $F_2=0$), the field strength $F_2$ (at constant $F_1$) and the phase shift $\Delta\phi=\phi_2-\phi_1$ on entrainment in a synchronized system ($K=5$). In particular, we map out the regions of the ($\Omega,f_1$), ($\Omega,F_2$) and ($\Omega,\Delta\phi$) parameter spaces occupied by each steady state (phase diagram), and perform a numerical bifurcation analysis of the transitions between entrained and disrupted states. Note that at zero external field ($F_1=F_2=0$) a fraction of the system's oscillators ($\rho_1=\rho_2=\sqrt{1-2/K}$) becomes synchronized if the coupling $K$ is larger than the critical value $K_C=2$. Thus, at the chosen coupling $K=5$, the oscillators in communities $G_1$ and $G_2$ remain strongly synchronized in a weak field, but increasing field heterogeneity can destroy this synchronization as will be shown below.


Following recent work on the transition between disrupted states in the homogeneous field case \cite{Wright2020}, the steady state of each group of oscillators $G_m$ is classified as entrained (E), rotating (R) or oscillating (O) based on the dynamics of the group's complex order parameter \(z_m\). In the homogeneous field case, the order parameter $z$ rotates at the field frequency with constant amplitude $\rho$ if the system is entrained, whereas in disrupted states, $z$ either oscillates or rotates periodically in the field's frame of reference. The entrained state corresponds to a stationary solution of Eqs.~(\ref{eq: rho1})--(\ref{eq: psi2}), i.e., an attractive fixed point of these equations. In disrupted states, the group amplitude $\rho$ oscillates periodically, as oscillators repeatedly fall in and out of alignment with each other. The qualitative difference between oscillations and rotations of $z$ is expressed in the dynamics of the group phase $\psi$. In the oscillating state, $\psi$ oscillates about a constant mean value, whereas in the rotating state, $\psi$ rotates continuously through $2\pi$. As a result, the average angular velocity is zero in the oscillating state and non-zero in the rotating state. In other words, oscillators are on average phase- and frequency-locked to the field in an oscillating state but drift relative to the field in a rotating state. The above-described dynamics of $z$ in disrupted states correspond to periodic orbits in the frame rotating at the field frequency $\sigma$. Notably, these orbits are topologically distinct. In the oscillating state, orbits lie outside the singular point $z=0$, but in the rotating state, orbits encircle the singular point. We may therefore characterize orbits by the number of counterclockwise turns (winding number $n$) around the singular point $z=0$ in a single period $T$, as demonstrated in \cite{Wright2020}. Oscillating states have winding number $n=0$, and rotating states winding number $n=\pm1$, depending on whether the field frequency is negative (plus) or positive (minus). To an observer in the laboratory (non-rotating) frame, the relationship between the dynamics (average angular velocity $v$) and topology (winding number $n$) of disrupted states is
\begin{equation}
  \label{eq:groupvel}
  v = \sigma + \frac{2\pi}{T}n,
\end{equation}
as proven in \cite{Wright2020}. This relationship shows	that, when the system transitions between an oscillating state ($n=0$) and a rotating state ($n=-1$) at some positive field frequency, the observer in the laboratory will witness an abrupt drop in
the average angular velocity.

Given the dependence of the global order parameter $Z$ on $z_m$ (see Eq.~(\ref{eq:zm})), we classify the overall steady state of the system using an ordered sequence of individual group states: EE if both groups are entrained, OO if both groups are oscillating, RR if both groups are rotating, and OR (RO) if $G_1$ is oscillating and $G_2$ is rotating ($G_1$ is rotating and $G_2$ is oscillating.).


\subsection{\label{sec:fracf} Fraction of field-exposed oscillators}

We begin by studying how field heterogeneity impacts the entrained state when only a fraction of synchronized oscillators $f_1$ is exposed to a periodic external field with varying strength $F_1$ and frequency $\sigma=\Omega$. To this end, we numerically solve Eqs.~(\ref{eq: rho1})-(\ref{eq: psi2}) for $F_2=0$ and $K=5$ while varying the field strength $F_1$ and the detuning parameter $\Omega$.

In the case of a uniform field, the group phase (the phase $\psi$ of the order parameter) is locked to the field phase ($\phi$) in the entrained state. As the field frequency is increased, and the system approaches the critical boundary where the entrained state is disrupted, the phase lag $\phi-\psi  > 0$ begins to increase. In the case of a heterogeneous field, when only a fraction  $f_1$ of oscillators in group $G_1$ are exposed to the field, and the remaining oscillators in group $G_2$ are not, both groups of oscillators are also phase-locked to the field phase $\phi_1$, but with distinct phase lags, such that  $\phi_1 > \psi_1 >\psi_2$, i.e., group $G_1$ follows the field and group $G_2$ follows group $G_1$. The phase lag $\delta \psi=\psi_1 - \psi_2$ increases with increasing detuning $\Omega$ or decreasing $f_1$ up to a critical value $\leq \pi/2$, at which point entrainment is disrupted.

\begin{figure}[htp]
\includegraphics[width=\linewidth]{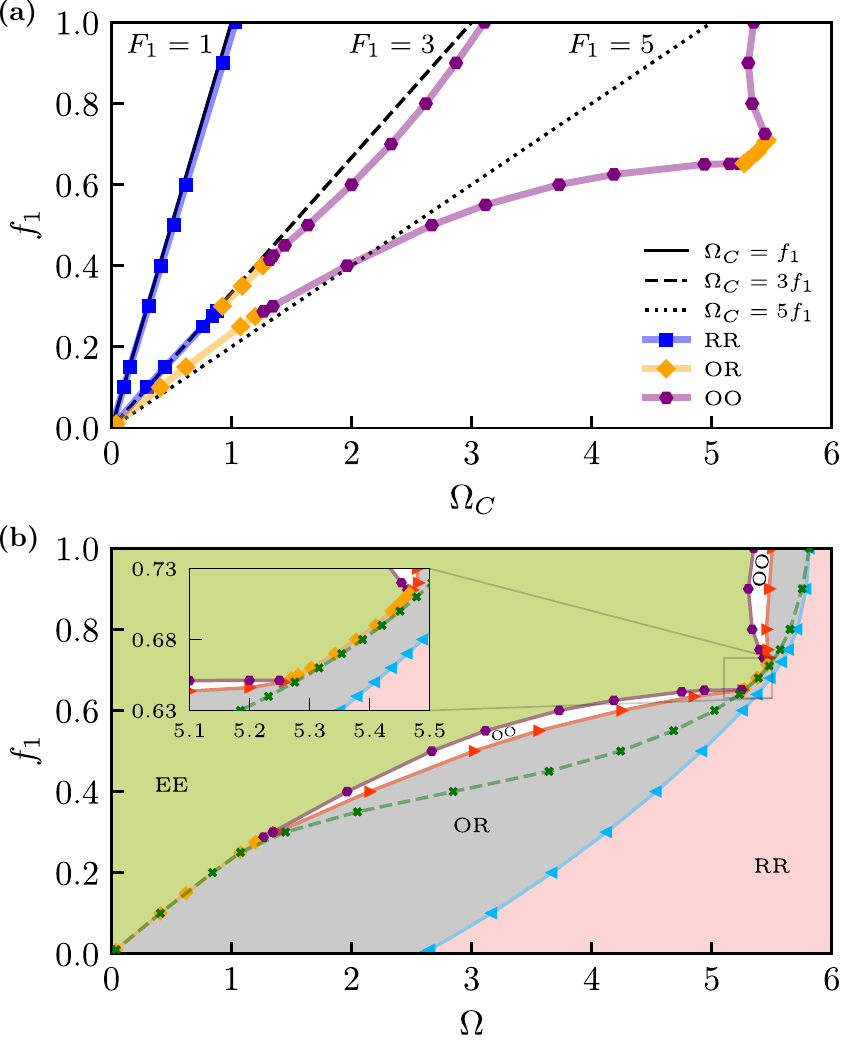}
	\caption{\label{fig:figure_2} Fraction of oscillators \(f_1\) subject to an external field with frequency \(\sigma=\Omega\) (detuning parameter) and strength \(F_1\), in a system of all-to-all interacting oscillators with coupling strength \(K=5\). The steady state of each group of oscillators is classified as entrained (E), rotating (R) or oscillating (O), and the overall steady state (or phase) of the system is described by the ordered sequence of individual group states. Panel (a) shows the critical detuning $\Omega_C$ at which entrainment is disrupted for a given fraction of field-exposed oscillators \(f_1\), and field strength \(F_1\) equal to 1, 3 and 5. Markers indicate the dynamical state of the system at the point where entrainment is disrupted, and the linear relationship for the critical detuning \(\Omega_C = f_1 F_1\) is presented for \(F_1\) equals 1 (full), 3 (dashed) and 5 (dotted). Panel (b) presents the nature and extent of different disrupted states (phase diagram) in the \((\Omega,f_1)\) parameter space at \(F_1=K=5\). Triangles indicate transitions where group $G_1$ (left-facing in blue) or group $G_2$ (right-facing in red) ceases to oscillate and begins to rotate. Green crosses indicate where the global order parameter $Z= f_1z_1 + f_2 z_2$ defined by Eq. (\ref{eq:zm}) transitions into a rotating state (R). As in panel (a), orange diamonds and purple hexagons indicate EE-OR and EE-OO transitions, respectively.}
\end{figure}

As shown in Fig.~\ref{fig:figure_2}(a), when coupling strength is sufficiently larger than field strength (see $F_1=1$ and $F_1=3$), decreasing the fraction of field-exposed oscillators causes entrainment to become disrupted at a lower critical detuning $\Omega_C$. From a physical point of view, this result can be qualitatively explained by the fact that decreasing the fraction of field-exposed oscillators makes it harder for these oscillators to entrain the remaining non-exposed oscillators. In particular, we note that when $F_1$ is sufficiently smaller than $K$, the critical detuning is directly proportional to $f_1$, as shown for $F_1=1$ in Fig.~\ref{fig:figure_2}(a). The latter dependence follows from Eqs.~(\ref{eq: psi1}) and (\ref{eq: psi2}), under the assumption that the entire system is strongly and equally synchronized (\(\rho_1 \simeq \rho_2 \simeq 1\)) close to the critical detuning if $F_1$ is sufficiently smaller than $K$, yielding \(\Omega_C\simeq f_1 F_1\). When coupling and field strengths become comparable, as shown for $F_1=5$ in Fig.~\ref{fig:figure_2}(a), exposing particular fractions of oscillators to the field (around $f_1=0.7$) can increase the critical detuning compared to the homogeneous field case ($f_1=1$).

The dynamics of disrupted states just above $\Omega_C$ are likewise dependent on the fraction of field-exposed oscillators and on the field strength. In the regime where the critical detuning is well-described by the linear relationship \(\Omega_C\simeq f_1 F_1\), exposed and non-exposed groups display rotating dynamics (see Section~\ref{sec:kmhf} for more details) when entrainment is disrupted, as shown in Fig.~\ref{fig:figure_2}(a) (navy squares). However, when the field strength increases (see $F_1=3$ and $F_1=5$), the dynamics of disrupted states vary with the fraction of field-exposed oscillators, and include states where the field-exposed group ($G_1$) is oscillating while the non-exposed group ($G_2$) is rotating (orange diamonds), or where both groups are oscillating (purple hexagons).

When $F_1=5$, Fig.~\ref{fig:figure_2}(b) shows that continuously increasing the detuning above $\Omega_C$ at fixed $f_1$ also causes transitions between disrupted states. Equivalently, decreasing $f_1$ at fixed detuning causes transitions between different disrupted states. In both cases, a larger change is required to cause $G_1$ (the subset of oscillators exposed to the field) to transition into a rotating state compared to $G_2$. If the detuning becomes sufficiently large, the system is found in an RR state for any $f_1$. Viewed together, the results in Fig.~\ref{fig:figure_2} show that, compared to the homogeneous field case ($f_1=1$) at the same field strength, the fraction of field-exposed oscillators changes both the critical detuning $\Omega_C$ at which entrainment is disrupted and the dynamics of disrupted states along the critical boundary.

\subsection{Heterogeneous field strength}
\label{sec:hetfs}

Next, we consider the impact of heterogeneous field strength by fixing the field strength $F_1=1$ acting on $G_1$, and simultaneously varying field strength $F_2$ acting on $G_2$ and the detuning $\Omega$. In order to control for the effect of field phase, we simply set both field phases $\phi_1=\phi_2=0$. In addition, since the effect of the fraction of oscillators in each group is to scale intra- and inter-group coupling strength, as can be inferred from Eqs.~(\ref{eq:rho_m}) and (\ref{eq:psi_m}), this effect was controlled for by setting $f_1=f_2=0.5$.

\begin{figure}[htp]
\includegraphics[width=\linewidth]{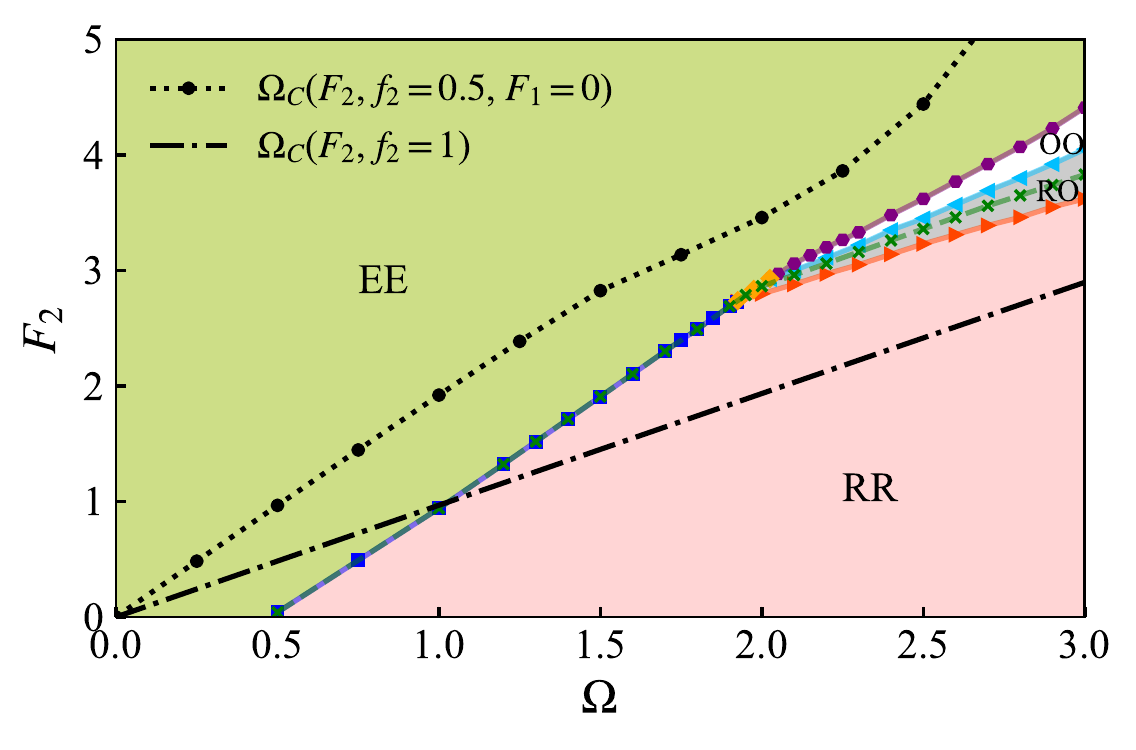}
\caption{\label{fig:figure_3} Phase diagram of ($\Omega$, $F_2$)-parameter space, for coupling strength $K=5$, and a fraction $f_1=0.5$ of oscillators exposed to local field strength $F_1=1$. Transition boundaries are indicated for EE-RR (navy squares), EE-RO (orange diamonds), and EE-OO (purple hexagons) transitions, as well as for the O-R transition in $G_1$ (left-facing triangles in blue) and $G_2$ (right-facing triangles in red). Green crosses indicate where the global order parameter $Z= (z_1 + z_2)/2$ defined by Eq. (\ref{eq:zm}) transitions into a rotating state (R). The dash-dotted line indicates how the critical detuning $\Omega_C$ for transitions between entrained and disrupted states varies with $F_2$ in the homogeneous field case (following \cite{Childs2008}), i.e., $f_2=1$ and $f_1=0$, (equivalently one can choose $f_2=0$, $f_1=1$ and vary $F_1$). The dotted line with circles indicates how  $\Omega_C$ varies with $F_2$ in the absence of $F_1$. }
\end{figure}

As shown in Fig.~\ref{fig:figure_3}, increasing $\Omega$ at constant $F_2$ causes the system to transition from an entrained state, where both groups are frequency- and phase-locked to the local field (EE) into different disrupted states (RR, RO and OO). The critical detuning $\Omega_C$ at which entrainment is disrupted was found to increase monotonically with $F_2\geq 0$. Compared to the homogeneous field case at the same local field strength (dash-dotted line), decreasing $F_2$ below $F_1=1$ increases $\Omega_C$, and increasing $F_2$ above $F_1=1$ decreases  $\Omega_C$. However, when compared to exposing only half of the system to $F_2$ (dotted line), applying a local field $F_1=1$ to the previously unexposed group of oscillators increases $\Omega_C$ for any $F_2$.

Increasing $F_2$ was also found to change the nature of disrupted states just above $\Omega_C$, from RR, through RO, to OO. At constant $F_2$, along the EE-OO transition, subsequent increases in $\Omega$ cause the system to first undergo an OO-RO transition, and then an RO-RR transition. Likewise, for the EE-RO transition, a subsequent increase in $\Omega$ causes the system to undergo an RO-RR transition. In other words, group $G_2$, which is exposed to the largest field strength ($F_2>F_1$), is the last to enter a rotating (drifting) state for increasing detuning, or decreasing local field strength.

These results show that field strength heterogeneity can both extend and decrease the range of field frequencies at which a system of coupled phase oscillators is disrupted, as well as change the nature of the disrupted states along the critical boundary.



\subsection{Heterogeneous field phase}
\label{sec:hetfp}

Lastly, we studied the effect of a phase shift (or time delay) $\Delta \phi=\phi_2-\phi_1$ in the local field phases experienced by each group for two distinct cases. In the first case, the local field strength acting on each group is equal: $F_1=F_2=3$. In the second case, $F_2>F_1$ ($F_2=3$ and $F_1=1$). For each case, we numerically solved Eqs.~(\ref{eq: rho1})-(\ref{eq: psi2}) for varying $\Delta \phi$ (setting $\phi_1=0$ and varying  $\phi_2$) and detuning $\Omega$, and controlled for the effect of group size by setting $f_1=f_2=0.5$.

\begin{figure*}[htp]
\includegraphics[scale=0.8]{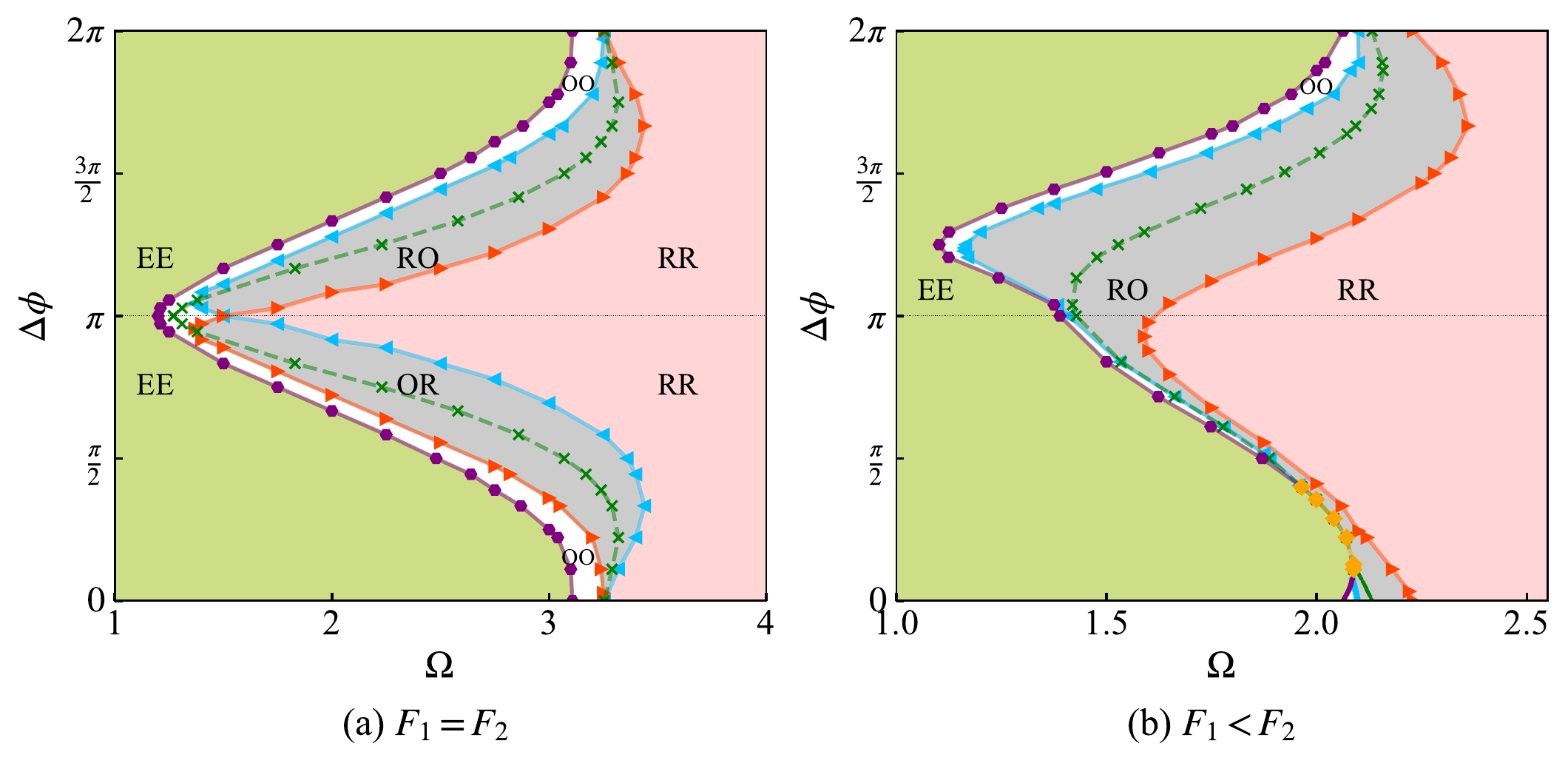}
\caption{\label{fig:figure_4} Phase diagram of ($\Omega$, $\Delta \phi$)-parameter space, for coupling strength $K=5$, and a fraction $f_1=0.5$ of oscillators exposed to local field strength $F_1$, for (a) $F_1=F_2=3$ and (b) $F_1<F_2$ ($F_1=1$, $F_2=3$). Transition boundaries are indicated for EE-OO (purple hexagons) and EE-RO (orange diamonds) transitions, and the O-R transition in $G_1$ (left-facing triangles in blue) and $G_2$ (right-facing triangles in red). Note that $G_1$ does not enter an oscillating state in panel (b), where the EE-OO transition is replaced with an EE-RO transition for $\pi/9<\Delta \phi< 2\pi/5$, and the O-R transition in $G_1$ is thus absent. Green crosses indicate where the global order parameter $Z= (z_1 + z_2)/2$ defined by Eq. (\ref{eq:zm}) transitions into a rotating state (R).}
\end{figure*}

The first case provides a useful means of comparison for the impact of $\Delta \phi$ relative to the homogeneous field case, where $\Delta \phi=0$. As shown in Fig.~\ref{fig:figure_4}(a), compared to the homogeneous field case, increasing $\Delta \phi$ decreases the critical detuning $\Omega_C$ at which entrainment is disrupted, until it becomes minimum at $\Delta \phi = \pi$.  In addition, $\Omega_C$ was found to vary symmetrically about $\Delta \phi =\pi$, i.e., independently of whether $\phi_2$ leads or lags $\phi_1=0$. Thus, if the detuning $\Omega$ is sufficiently small, the system remains in an EE state for any $\Delta \phi$. Similarly,  Fig.~\ref{fig:figure_4}(a) shows that if $\Omega$ is large enough, the system remains in an RR state for any $\Delta \phi$. For intermediate $\Omega$, continuously varying $\Delta \phi$ disrupts entrainment and causes transitions between different disrupted states. Interestingly, we note that OR and RO states are determined by the leading local field phase: in OR states, $\phi_2$ leads $\phi_1$, and in RO states, $\phi_1$ leads $\phi_2$. So, for example, at $\Omega=3$, continuously increasing $\Delta \phi=\phi_2-\phi_1$ from $0$ to $2\pi$ causes the system to transition through the following sequence of dynamical states: EE-OO-OR-RR-RO-OO-EE.

In the second case, we consider the effect of a difference in the local field phases on a system where $F_2 > F_1$ (for $F_1=1$ and $F_2=3$). In the presence of this heterogeneity in the local field strength, Fig.~\ref{fig:figure_4}(b) shows that increasing $\Delta \phi $ from approximately $\pi/9$ to $5\pi/4$ also decreases the critical detuning $\Omega_C$ at which entrainment is disrupted, until $\Omega_C$  becomes minimum at $\Delta \phi \simeq 5\pi/4$. A similar trend of decreasing $\Omega_C$ is observed if $\Delta \phi$ is decreased from $2\pi$ to $\simeq 5\pi/4$, i.e., if $\phi_2$ lags $\phi_1$ by up to $\pi/2$, or leads $\phi_1$ by up to $\pi/4$. However, increasing $\Delta \phi$ between  $0$ and $\simeq \pi/9$ increases $\Omega_C$.  Compared to the $F_1=F_2$ case, depicted in Fig.~\ref{fig:figure_4}(a), we see that $\Omega_C$ no longer varies symmetrically with $\Delta \phi$. In addition, Fig.~\ref{fig:figure_4}(b) shows that entrainment can now be disrupted directly into an RO state (orange diamonds) as well as into an OO state (purple hexagons). For sufficiently small or large $\Omega$, varying $\Delta \phi$ causes no change in the system, which remains in an EE or RR state, respectively. For intermediate $\Omega$, varying $\Delta \phi$ causes transitions between disrupted states. However, we note that the OR state is no longer accessible at any $\Omega$ or $\Delta\phi$, i.e., the group exposed to the smaller field does not display oscillating dynamics.

These results show that the Kuramoto model in a heterogeneous field is very sensitive to differences between the local field phases acting on distinct groups of oscillators. Within a range of field frequencies, a delay or shift between local field phases can both decrease and increase the critical detuning at which entrainment is disrupted, when compared to the homogeneous field case. However, an increase in the critical detuning was only observed for a very narrow range of phase shifts in the presence of field strength heterogeneity. In the absence of field strength heterogeneity, equal clockwise and counterclockwise phase shifts were found to alter the critical boundary with disrupted states symmetrically, without affecting the nature of the disrupted state along the critical boundary with entrained states.

\subsection{Stability diagram}
\label{sec:stab}

\begin{figure}[htp]
\includegraphics[width=\linewidth]{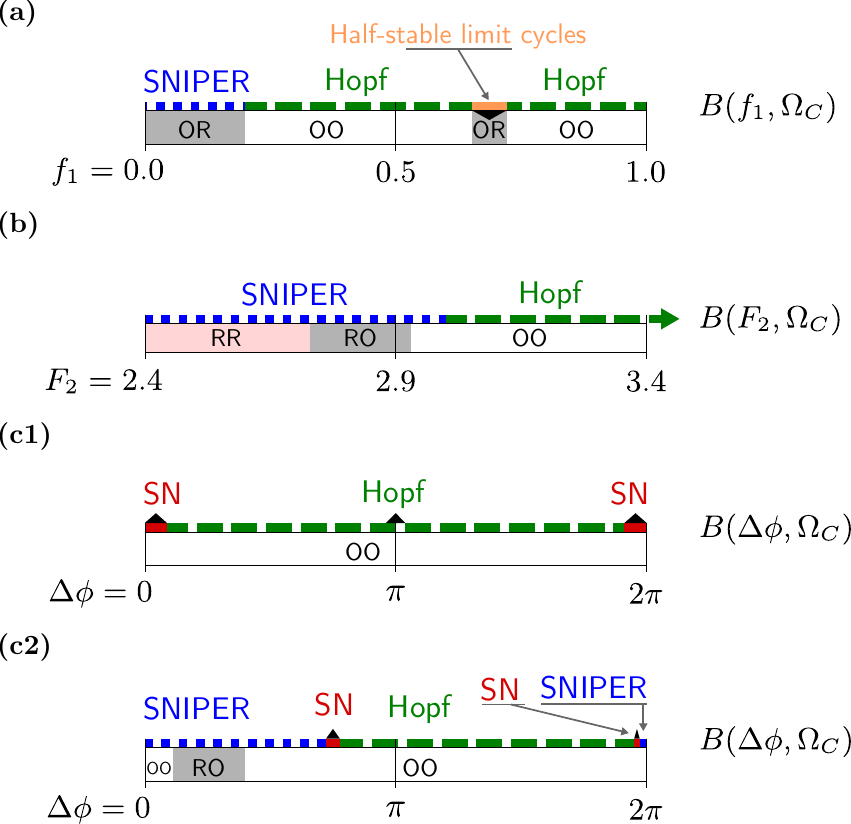}
\caption{\label{fig:figure_5} Stability diagram along the boundary between entrained and disrupted states. The boundary $B$ is defined by the critical detuning $\Omega_C$ and the fraction of field-exposed oscillators  $f_1$  (a), the field strength $F_2$ (b), and the field phase shift $\Delta\phi$ for homogeneous (c1) and heterogeneous (c2) field strength, from the corresponding phase diagrams Figs. \ref{fig:figure_2}(b), \ref{fig:figure_3}, and \ref{fig:figure_4}(a) and \ref{fig:figure_4}(b), respectively. Black triangles indicate the existence of bistable regions along the critical boundary, where the steady state of the system depends on the initial conditions $(\rho_m,\psi_m)$ of each group $G_m$ (see text for details). Downward-facing triangles (black) in  panel (a) indicate bistable regions where entrained and disrupted states coexist. Upright triangles (black) in panels (c1) and (c2) indicate bistable regions where only entrained states exist. Saddle-Node (SN), Hopf and SNIPER bifurcations are indicated in function of $f_1$, $F_2$ or $\Delta\phi$ by full red, dashed green and dotted blue lines, respectively.}
\end{figure}

Based on a numerical bifurcation analysis of Eqs.~(\ref{eq: rho1})-(\ref{eq: psi2}), we identified the transition mechanisms along the critical boundary between entrained and disrupted states, which are summarized in the stability diagram presented in Fig.~\ref{fig:figure_5}. These results show that field heterogeneity plays an important role in determining the mechanisms that disrupt entrainment in a system of coupled phase oscillators, in addition to the dynamics of disrupted states. The identified mechanisms include some of the bifurcations reported for the homogeneous field case \cite{Childs2008}, namely saddle-node (SN), saddle node infinite period (SNIPER) and Hopf bifurcations. Similarly to the homogeneous case, disrupted states are not uniquely associated with a particular bifurcation, as shown in Fig.~\ref{fig:figure_5}. For example, the EE-OO transition can take place through SN, SNIPER and Hopf bifurcations. However, our analysis also revealed the presence of half-stable limit cycles and other bistable regions.

Figure~\ref{fig:figure_5}(a) shows that, for a critical field frequency (detuning $\Omega_C$), the fraction of field-exposed oscillators simultaneously determines the disruption mechanisms and the dynamics of disrupted states. When $F_2$ becomes non-zero, as shown in Fig.~\ref{fig:figure_5}(b), continuously increasing $F_2$ (at $F_1=1$) can change the bifurcation mechanism along the critical boundary while leaving the nature of the disrupted state unchanged. For example, an EE-OO transition can occur either through a SNIPER or a Hopf bifurcation for an infinitesimal change in $F_2$. Conversely, an infinitesimal change in $F_2$ can also cause a transition into distinct disrupted states through the same bifurcation, as in the case of EE-RO and EE-OO transitions under a SNIPER bifurcation. Introducing further field heterogeneity in the form of a finite phase shift $\Delta \phi$ can also alter the bifurcation underlying the EE-OO transition, e.g. from a SNIPER, through a SN, to a Hopf bifurcation, as can be seen by comparing Figs.~\ref{fig:figure_5}(b) and (c2) (at $F_2=3$). Lastly, we note that when $F_1=F_2=3$, the EE-OO transition can also take place through a SN bifurcation in a small range of $|\Delta\phi|>0$, which is symmetric with respect to a change in the leading phase, as shown in Fig.~\ref{fig:figure_5}(c1).

The phase diagrams presented in the preceding sections were obtained from initial conditions $z_1(t_0)=z_2(t_0)=1$ at $t_0=0$. However, our bifurcation analysis also revealed different bistable regions along the critical boundary between entrained and disrupted states, as shown in Fig.~\ref{fig:figure_5} (black triangles). The bistable section of the $(\Omega_C,f_1)$ boundary presented in Fig.~\ref{fig:figure_5}(a) (downward-facing triangles) spans from $f_1$ approximately between $0.65$ and $0.71$, where the system's only fixed point is stable, and remains stable for a range of $\Omega\geq\Omega_C$. In other words, for $f_1$ between $0.65$ and $0.71$, the EE-OR transition is not determined by a change in the stability of the single fixed point $\tilde{Z}= f_1 \tilde{z}_1+f_2 \tilde{z}_2$. Within the same region of the $(\Omega,f_1)$ parameter-space, we identified two limit cycles: limit cycle $c_1$, encircling $\tilde{z}_1=\tilde{\rho}_1e^{i\tilde{\psi}_1}$, and limit cycle $c_2$, encircling $\tilde{z}_2=\tilde{\rho}_2e^{i\tilde{\psi}_2}$. Solving Eqs.~(\ref{eq: rho1})--(\ref{eq: psi2}) for different initial conditions $Z(t_0)=f_1 z_1(t_0) +f_2 z_2(t_0)$ at $t_0=0$, we found that limit cycles $c_1$ and $c_2$ are half-stable: unstable if $z_1(t_0)$ is located within $c_1$ and $z_2(t_0)$ is located within $c_2$, and stable otherwise. In the unstable case, all orbits starting at $z_1(t_0)$  simply spiral into $\tilde{z}_1$, and orbits starting at $z_2(t_0)$ simply spiral into $\tilde{z}_2$, so that $G_1$ and $G_2$ become entrained. Bistable regions were also identified along the $(\Omega_C,\Delta \phi)$ boundary in Figs.~\ref{fig:figure_5}(c1) and (c2) (upright triangles). The bistable sections along the boundary in Fig.~\ref{fig:figure_5}(c1) are both symmetric with respect to  $\Delta \phi$, i.e., with respect to the replacement $\phi \rightarrow 2\pi-\phi$. The section defined by a range of $\Delta \phi$ around $\pi$ (Hopf bifurcation) has two stable fixed points (symmetrically located about the origin in the first and third quadrants), and the section defined by the range of $\Delta \phi$ around $0$ or $2\pi$ (SN bifurcation) has one stable fixed point and one stable limit cycle. Compared to the $F_1=F_2=3$ case of Fig.~\ref{fig:figure_5}(c1), Fig.~\ref{fig:figure_5}(c2) shows that decreasing $F_1=1$ splits the bistable section centered around $\Delta \phi = 0$ (or 2$\pi$), shifting each new section towards $\Delta \phi=\pi$, without changing the nature of the transition at the critical boundary or the bistability itself. The new bistable sections border a SN bifurcation, and contain one stable fixed point and one stable limit cycle. At the $(\Omega_C,\Delta \phi)$  boundary, all stable fixed points either disappear or become unstable. These results show that a careful choice of the fraction of field-exposed oscillators or the phase shift between local phases can determine whether a system will be entrained or disrupted, depending on its state when a heterogeneous field is applied at specific frequencies.

\section{Dynamics of disrupted states}
\label{sec:dynamics}

In this section we discuss the dynamics of the different disrupted states found in a system of uniformly coupled phase oscillators exposed to field strength heterogeneity, for the particular case considered in Sec.~\ref{sec:hetfs}, where the field splits the system into two interacting groups of oscillators. Our focus is on the main qualitative differences between states, and their dependence on the field frequency, from the point of view of an observer in the laboratory (non-rotating frame of reference). In particular, we consider how the real part of the complex order parameter $\mathrm{Re}(z_m)=\rho_m \cos(\psi_m +\sigma t)$ varies over time, and the corresponding spectral density $S(\omega)$.

\begin{figure}[H]
\centering
\includegraphics[width=\columnwidth]{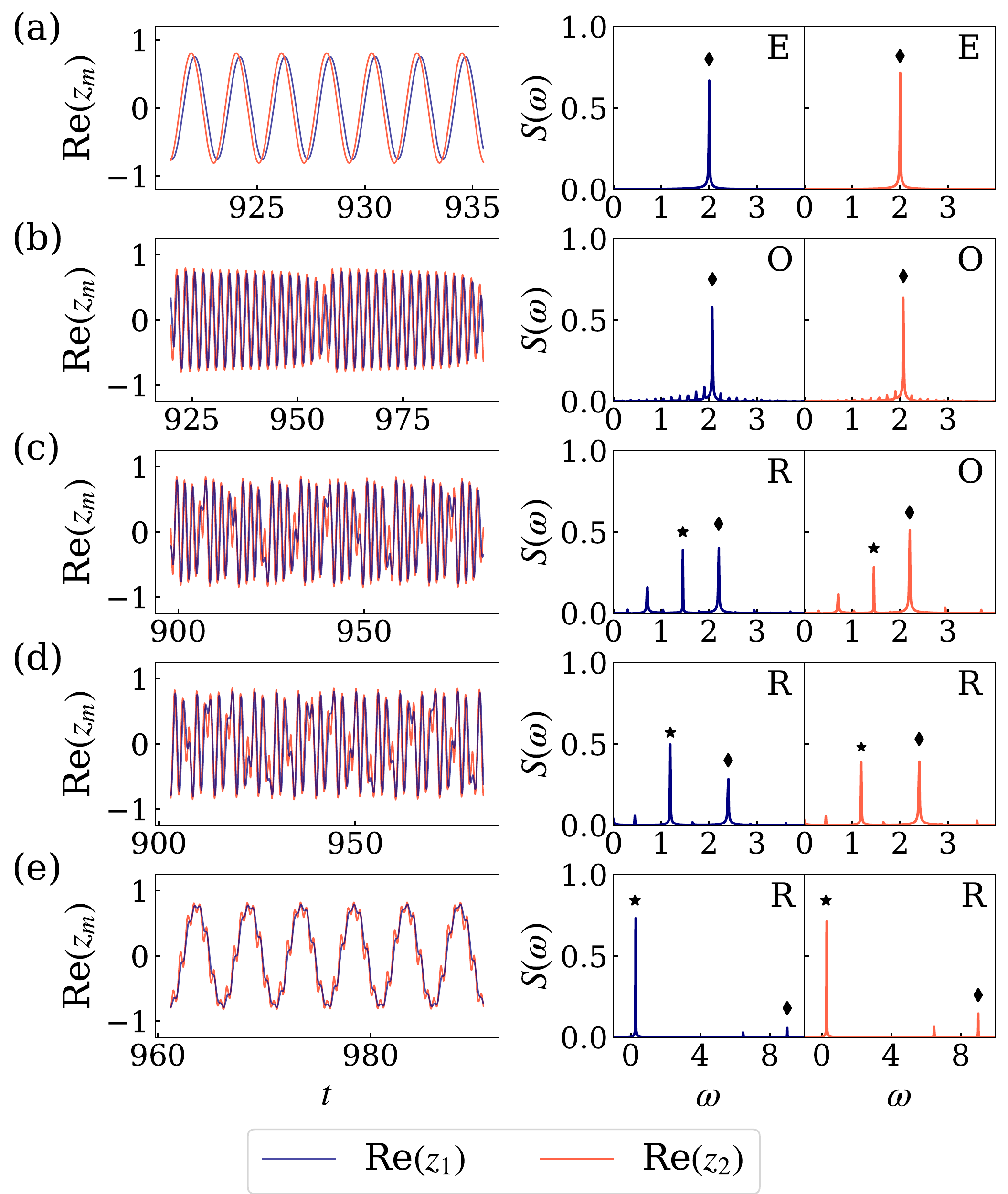}
\caption{\label{fig:figure_6} Dynamics of two identical groups ($G_1$ and $G_2$) of coupled Kuramoto oscillators subject to heterogeneous field strengths ($F_1=1$ and $F_2=3$), at field frequency $\sigma$, in the stationary (laboratory) frame of reference. Horizontal panels (a)-(e) show the time-dependent behavior of $\mathrm{Re}(z_m)=\rho_m \cos\left(\psi_m+\sigma t\right)$, and the corresponding spectral density $S(\omega)$, for group $G_{1}$ (navy) and $G_{2}$ (orange), at field frequency (a) $\sigma=2.0$ (EE), (b) $\sigma=2.07$ (OO), (c) $\sigma=2.2$ (RO), (d) $\sigma=2.4$ (RR), and (e) $\sigma=9.0$ (RR). The steady state of each group is classified as entrained (E), oscillating (O), or rotating (R), based on the dynamics of the complex order parameter $z_m$, as discussed in the introduction to Sec.~\ref{sec:results}. Diamonds in the spectral density indicate the field frequency, and stars indicate the average group velocity of groups in the rotating state (R). Note that panel (e) corresponds to the case where the field frequency is much larger than required to disrupt entrainment. For reference, each panel in this figure corresponds to a single point $(\Omega=\sigma,F_2=3)$ in Fig.~\ref{fig:figure_3} of Sec~\ref{sec:hetfs}.
}
\end{figure}

As shown in Fig.~\ref{fig:figure_6}, the dynamics of the group amplitude $\rho_m$ (extent of synchronization) and phase $\psi_m$ in different dynamical states are reflected in the time-varying signal $\mathrm{Re}(z_m)$ and its spectral density $S(\omega)$. In general, regardless of the dynamical state, we found that the frequency components revealed by $S(\omega)$ are identical in both groups, which we attribute to the coupling between groups. At sufficiently low field frequencies, the system is entrained and both groups of oscillators are frequency-locked to the field, so that $\mathrm{Re}(z_m)$ oscillates periodically at the field frequency (EE state), as typified in Fig.~\ref{fig:figure_6}(a). When both groups enter an oscillating state (OO state), $S(\omega)$ reveals additional frequency components, as shown in Fig.~\ref{fig:figure_6}(b), but the field frequency $\sigma$ (diamond) remains dominant. However, when the first group enters the rotating state (i.e., the system enters the RO state), the spectral density is characterized by a drop in the field frequency component (relative to the OO state), and the presence of smaller frequencies with significant spectral densities, as shown in Fig.~\ref{fig:figure_6}(c). In particular, the spectral density contains a significant frequency component to the left of the field frequency $\sigma$ (diamond), at the average angular velocity of the rotating group $v$ (star). The appearance of this second frequency component is in agreement with Eq.~(\ref{eq:groupvel}), which shows that the frequency difference $\sigma - v = 2\pi/T$, where $T$ is the group's period of rotation at positive field frequency (see the introduction to Sec.~\ref{sec:results} for further detail). Figures~\ref{fig:figure_6}(d) and (e) show that subsequent increases in the field frequency cause an increase in the period $T$, as the corresponding component $v$ is shifted towards smaller frequencies, and becomes increasingly dominant in the spectral density.

\begin{figure}[H]
\centering
\includegraphics[width=0.95\columnwidth]{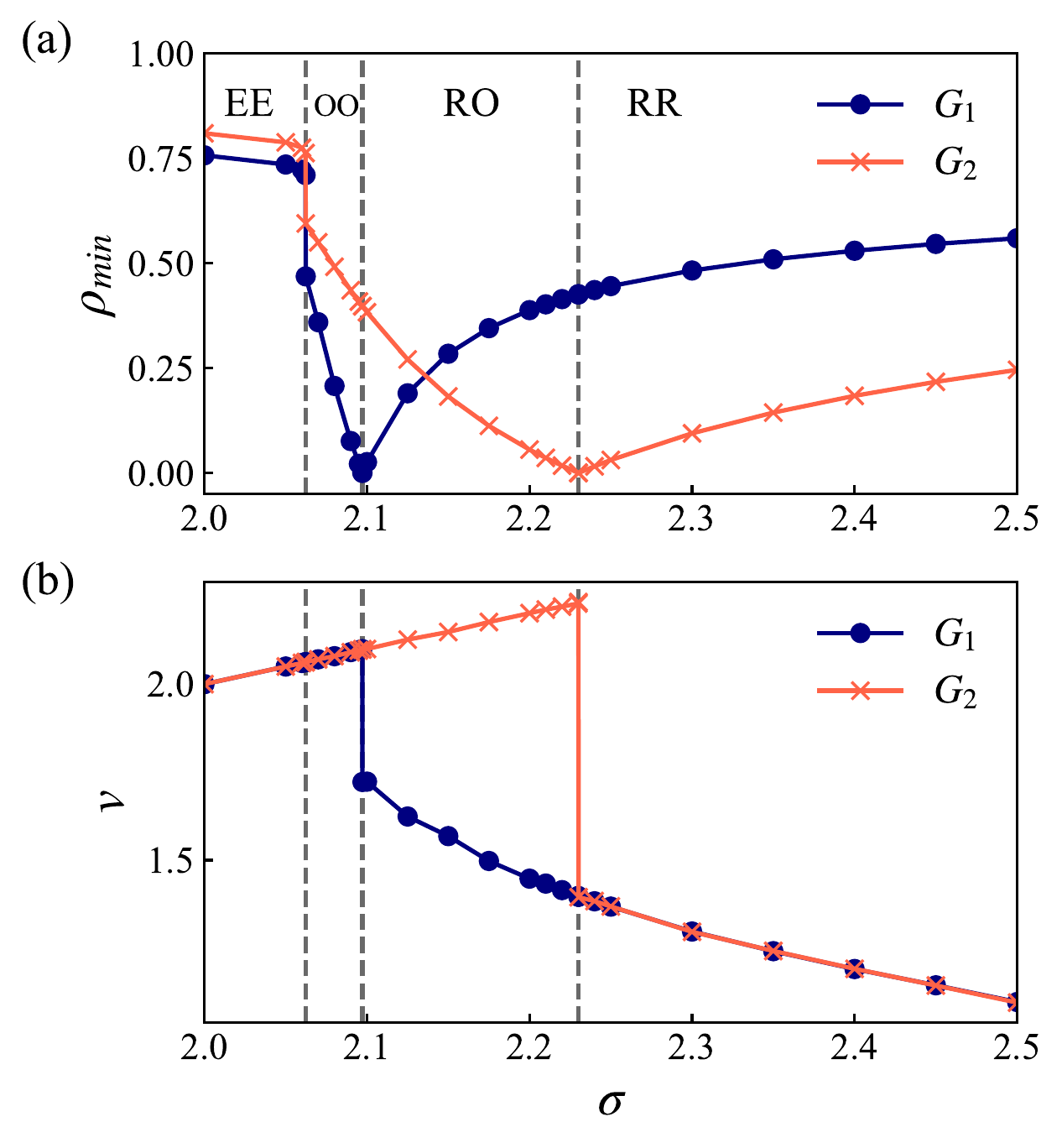}
	\caption{\label{fig:figure_7} (a) Minimum group amplitude $\rho_{min}$ and (b) average group angular velocity $v$ in the stationary (laboratory) frame of reference, for two identical groups $G_1$ (navy circles) and $G_2$ (orange crosses) of coupled Kuramoto oscillators subject to heterogeneous field strengths ($F_1=1$ and $F_2=3$), as a function of the field frequency $\sigma$. Dashed vertical lines indicate the critical field frequencies at which the system undergoes a transition.}
\end{figure}

Lastly, and identically to the homogeneous field case, the transition into a rotating state (R) under increasing field frequency is clearly reflected in the extent of synchronization within the group entering the rotating state. As previously discussed, disrupted states are characterized by an oscillation in the group amplitude $\rho_m$, which corresponds to individual oscillators falling in and out of alignment. As shown in Fig.~\ref{fig:figure_7}(a), the minimum extent of synchronization $\rho_{min} = \mathrm{min}(\rho_m)$ within each group varies with field frequency, and is minimum at the critical field frequency for the transition into rotating dynamics. Figure~\ref{fig:figure_7}(b) shows that the critical frequency is also characterized by an abrupt drop in the average group angular velocity $v$, identically to the homogeneous field case. Below the critical frequency, $v$ is equal to and grows linearly with the field frequency. Above the critical field frequency, $v$ decreases monotonically with the field frequency.


\section{Discussion}
\label{sec:discussion}

In this article, we demonstrated that a heterogeneous external field partitions a system of Kuramoto oscillators into several groups by acting on each group with a distinct local phase and strength. When the natural frequencies of oscillators are distributed according to a Lorentzian distribution, the set of dynamical equations for single oscillators is reduced to the set of explicit equations for the group amplitude and the group phase, as presented in Eqs. (\ref{eq:rho_m}) and (\ref{eq:psi_m}). Using the reduced equations, we studied the impact of field heterogeneity on entrainment in a system with two distinct groups of oscillators. Numerical analysis of the explicit dynamical equations for the Kuramoto model in a heterogeneous field showed that field heterogeneity can significantly alter the critical detuning at which entrainment is disrupted, the dynamics of disrupted states, and the mechanisms underlying the transition between entrained and disrupted states. In particular, we studied the impact of exposing only a fraction of oscillators to the field, varying the field strength on one group while the other remains constant, and introducing a phase shift between local field phases.

Viewed together, our results show that the heterogeneous field can both increase or decrease the range of field frequencies over which the system remains entrained when compared to the homogeneous field case. At the critical field frequency, phase- and frequency-locked synchronization between groups is broken, causing each group to enter a disrupted state. In this disrupted state, field heterogeneity determines the steady state dynamics of each group, which can differ between groups. The main difference between disrupted states concerns the dynamics of the group phase, which can either oscillate (oscillating state) or continuously drift (rotating state) in relation to the field phase. On average, the oscillating group phase follows the field at the field frequency, while the drifting phase lags behind at a smaller frequency. This additional frequency produced by the rotating group is identifiable in the spectral density of the corresponding order parameter $\mathrm{Re}\left(z_m\right)$ measured in the laboratory frame, and was shown to decrease with the field frequency (see Figs.~6(c)--(d) and Fig.~7).

Our numerical stability analysis revealed that many of the mechanisms underlying the transition between entrained and disrupted states are identical to the homogenous field case, namely saddle-node (SN), saddle node infinite period (SNIPER) and Hopf bifurcations, as shown in Fig.~\ref{fig:figure_5}. In addition, we also identified bistable regions of field heterogeneity, where the steady state of the system depends on the extent of synchronization and the phase of each group when the field is applied (initial conditions). These bistable regions include half-stable limit cycles for particular fractions of field-exposed oscillators at large local field strength, which have not been reported for the homogeneous field case.

The reduced dynamical equations presented in Eqs.~(\ref{eq:rho_m}) and (\ref{eq:psi_m}) remarkably simplify the study of entrainment in heterogeneous fields. We believe that the methodology employed in this work is highly relevant and easily applicable to the study of entrainment in real systems of synchronized phase oscillators, such as the brain's suprachiasmatic nucleus (SCN) and the related study of circadian rhythms. Although the reduced equations presuppose a Lorentzian distribution of natural frequencies among individual oscillators, we note that the Lorentzian distribution is similar to the Gaussian distribution employed in models of SCN dynamics \cite{Taylor2017}, insofar as both distributions are unimodal. In other words, our reduced equations describe a system where most oscillators have a natural frequency equal to the system average, and the number of oscillators with larger or smaller than average natural frequencies is symmetric and monotonically decreasing. Furthermore, our results for uniformly coupled oscillators clearly show that entrainment is dependent on the difference between local field phases, such as may be introduced by cues acting on different groups of SCN oscillators with a time delay. For example, one may consider the case where a photic cue acting on the SCN is not compatible with feeding times, causing circadian rhythms to become disrupted \cite{Mendoza2005,Mendoza2007,Heyde_2019}. Moreover, our findings concerning the dynamics of rotating states are compatible with the known dissociation between SCN circadian rhythms under light-dark cycles shorter than 24 hours, characterized by the appearance of a non-entrained rhythm with a period shorter than the light-dark cycle \cite{Campuzano1998}. Finally, we note that the Lorentzian distribution of natural frequencies (intrinsic periods) can in general differ between groups (different mean and spread), and intra- and inter-group couplings may also differ in general, similarly to what is observed in the core and the shell of the SCN \cite{Taylor2017}.

\begin{acknowledgments}
This work is funded by national funds (OE), through Portugal's FCT Funda\c{c}\~{a}o para a Ci\^{e}ncia e Tecnologia, I.P., within the scope of the framework contract foreseen in paragraphs 4, 5 and 6 of article 23, of Decree-Law 57/2016, of August 29, and amended by Law 57/2017, of July 19. E. A. P. W. acknowledges the financial support provided by FCT under PhD grant SFRH/BD/121331/2016.
\end{acknowledgments}

\bibliography{bibliography}
\end{document}